\documentclass[prb,showpacs,twocolumn,amsmath,amssymb,floatfix]{revtex4}
\usepackage{graphicx}
\usepackage{subfigure}
\usepackage{dcolumn}
\usepackage{epsfig}
\usepackage{amssymb}
\usepackage{amsmath}
\usepackage{rotating}
\usepackage{color}

\begin{document} 

\title{Making a splash with water repellency}
 
\author{Cyril Duez$^\dag$, Christophe Ybert$^\dag$, Christophe Clanet$^\ddag$, Lyd\'eric Bocquet$^\dag$}
\email{lyderic.bocquet@univ-lyon1.fr}
\address{$^\dag$ Laboratoire PMCN, Universit\'e Lyon 1, UMR CNRS 5586, 69622 Villeurbanne, France \\
$^\ddag$ IRPHE, UMR CNRS 6594, 13384 Marseille, France}
\maketitle

\begin{figure}[t]
\begin{center}
\includegraphics[width=8cm,height=!]{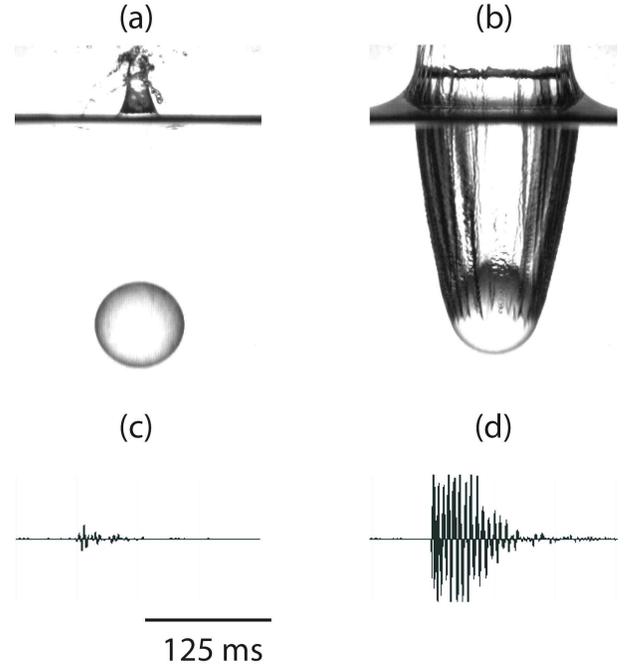}
\caption{{\it Top}: Photographs of the impact of two spheres differing only in wettability via a nanometric coating on their surface: (a) impact of a perfectly wetting sphere, with static contact angle $\theta_0\simeq15^\circ$; (b)
impact of a hydrophobic sphere with static contact angle $\theta_0\simeq100^\circ$. The impact velocity was 5.0 m.s$^{-1}$ in 
both cases,
corresponding to a $1.25$m height drop. 
Pictures were taken 15.5ms (left) and 15.0ms (right) after impact beginning. 
{\it Bottom}:
Time-dependent audio recordings of the impacts, as measured 
by a microphone $\sim 10$ cm from the impact point, for (c) a hydrophilic 
and (d) a hydrophobic sphere. The signal is 
% MODIF SUPP
{proportional to} 
%END MODIF SUPP
the acoustic pressure emitted during the impact. Units on the vertical scale are arbitrary (but identical). A big 'splash' is evident
for the hydrophobic sphere, while a tiny 'plop' is heard for the hydrophilic one. The sound is associated
with the rapid closure of the cavity (not shown).
\label{fig0}}
\end{center} 
\end{figure}

{\bf 
A 'splash' is usually heard when a solid body enters water at large velocity.
This phenomena originates from the formation of an air cavity resulting from the complex transient dynamics of the free interface during 
the impact.
The classical picture of impacts on free surfaces relies solely on fluid inertia, arguing that surface
properties and viscous effects are negligible at sufficiently large velocities. 
In strong contrast to this large-scale hydrodynamic viewpoint, we demonstrate in this study
that  the wettability 
of the impacting body is a key factor in determining the degree of splashing.  
This unexpected result is illustrated in Fig.\ref{fig0}: a large cavity is evident for an
impacting hydrophobic sphere (\ref{fig0}.b), contrasting with the hydrophilic sphere's impact under the very same conditions (\ref{fig0}.a).
This unforeseen fact is furthermore embodied in the dependence of
the threshold velocity for air entrainment on the contact angle of the impacting body, 
as well as on the ratio between the surface tension and fluid viscosity, thereby defining a critical capillary velocity. As a paradigm, we show that superhydrophobic impacters make a big 'splash'
for any impact velocity.
This novel understanding provides a new perspective for impacts on free surfaces, and
reveals that modifications of the detailed nature of the surface -- involving physico-chemical aspects at the nanometric scales -- provide an efficient and versatile strategy for controlling 
the water entry of solid bodies at high velocity.
}

The first systematic study of splashes was published more than one century ago by Worthington
\cite{Worthington}, who used high speed photography to examine impacts of drops
and solid bodies on a liquid surface.  
In the recent years, there has been a resurgence in interest in the physics of impact,
thanks in particular to the development of rapid video imaging. 
And new perspectives have emerged, showing that unforeseen mechanisms 
play a central role in impact: to cite a few, the inhibition of droplet rebound by adding tiny amounts of polymers \cite{vovelle}, the complex deformation dynamics of a rebounding drop 
\cite{Richard}, and the unexpected role of ambiant air on drop splashing  \cite{Nagel,Quere}.
%\NOTE{citer : air entrainment by viscous jets in a liquid bath \cite{Lorenceau} ?}
We consider in this study the situation of a solid body impacting a gas-liquid interface. This situation is obviously relevant for many naval applications, such as ship slamming and
air to sea weapons, and for any industrial coating process which involves the dipping of a solid object in a liquid bath (where air entrainement is to be avoided). 
The traditional description of an impact of a solid body on a free interface follows the footsteps of
von Karman and Wagner \cite{Karman,Wagner}, 
in which viscosity, 
surface tension and compressibility effects are neglected \cite{oliver,Howison}.
This idealized framework is formally justified by the fact that in the situations relevant
to impacts, the Reynolds ${\cal R}e$ and Weber $We$ 
numbers, quantifying the role of inertia versus respectively viscous and capillary effects, are very large. 
This is precisely the regime of interest in this study:
${\cal R}e=\rho Ua/\mu_L \gtrsim 10^4-10^5$ and $We=\rho U^2 a/\gamma_{LV} \gtrsimÊ10^3-10^4$
(for impacting body diameter $a$, velocity $U$, liquid density $\rho$, liquid viscosity $\mu_L$, and 
liquid-vapour surface tension $\gamma_{LV}$).
Accordingly, capillarity and viscosity are not expected to play any role in the impact and can be
ignored in this description.

Our experimental results contrast with this simple picture.
As illustrated in Fig.\ref{fig0}, two spheres that only differ by a nanometric coating which modifies 
wettability exhibit very different impact behavior: 
a huge air cavity is entrained for the hydrophobic sphere, while no such behavior
is observed fo the hydrophilic one.
However, apart from the static contact angle ($\theta_0\simeq 15^\circ$ versus $\theta_0\simeq100^\circ$), the
spheres are identical in terms of bulk material (glass), diameter, very low surface roughness and impact velocity ($U=5$m.s$^{-1}$). 
Moreover, during the experiment, a 'splash' is heard for the hydrophobic sphere, while only a tiny
'plop' is produced by the hydrophilic one, as shown in Fig.\ref{fig0} (c-d).
%This difference of sounds is shown qualitatively in 
%Fig.\ref{fig0}, where we show an audio recording measured by a microphone close to the impact point. 
\begin{figure}[t]
\begin{center}
\includegraphics[width=!,height=!]{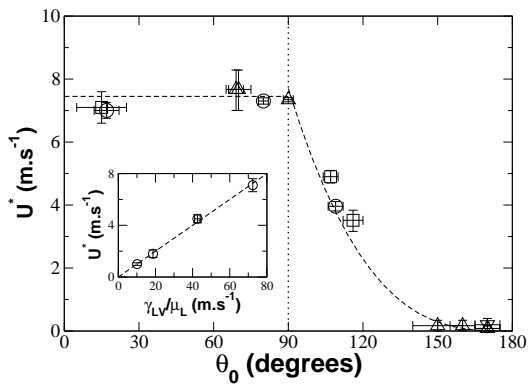}
\caption{Threshold velocity $U^{\star}$ for air entrainment as a function of (advancing) static contact angle $\theta_0$ 
of the impacting body. The dashed lines are the theoretical predictions based on relations (\ref{Unonwet}) and (\ref{Uwet}). 
The different symbols correspond to different bead diameters: 
$\square$ 25.4 mm (glass); $\bigtriangledown$ 20mm (aluminium); $\bigcirc$ 15mm (glass, steel); $\bigtriangleup$ 7mm (aluminium, steel). The beads are covered with various coatings to modify their
wettability (see Method section). To focus on wettability as the only surface parameter, only smooth objects have been considered in the present study (see Method section).
{\it Inset}: Dependence of the threshold velocity for a wetting glass sphere (25.4mm) 
on the ratio $\gamma_{LV}/\mu_L$. 
We used
various liquids to explore this dependence: water, isopropanol, ethanol, a 
water-glycerol mixture (20wt\% of glycerol).  
For these
fluids, the contact angle on the sphere surface was always below 10$^\circ$.
The dashed line is a linear prediction $U^\star=\xi \gamma_{LV}/\mu_L$ (with $\xi\simeq 0.1$).
%END MODIF SUPP
%A careful procedure has been followed to avoid pollution of the surfaces before impacts (see Method section). This latter point is crucial to control properly the state of the impacter's surface.
\label{fig2}}
\end{center} 
\end{figure}
This observation raises puzzling questions: how can a nanometric coating modify large scale hydrodynamics? 
More generally, how might capillarity affect the flow pattern in the limit of large Weber numbers?
To answer these questions, we have first explored the conditions required to create an air cavity, as illustrated 
in the above example.
By varying the velocity of the impacting body, we have demonstrated that an air cavity is created
during an impact only above a threshold velocity, $U^\star$, typically of a few meters per second. 
Furthermore this threshold velocity is found to depend on the (advancing) contact angle $\theta_0$ of the impacting body.  The experimental results for  $U^\star$  in water
are gathered in Fig.\ref{fig2} for spheres with various wettabilities. 
Going further, we measured the dependence of the threshold velocity $U^\star$ on the liquid
properties, by considering impacts on various liquids (with different viscosities and surface tensions)
for fixed wettability. 
As shown in the inset of Fig.\ref{fig2}, we found that $U^\star$ is proportional
to the capillary velocity, defined as $\gamma_{LV}/\mu_L$. 
To complete this exploration, we 
verified that the diameter of the impacting sphere does not influence the threshold (see Fig.\ref{fig2}),
neither gas pressure (varied between 0.1 and 1 atm).
\begin{figure}[t]
\begin{center}
\includegraphics[width=8cm,height=!]{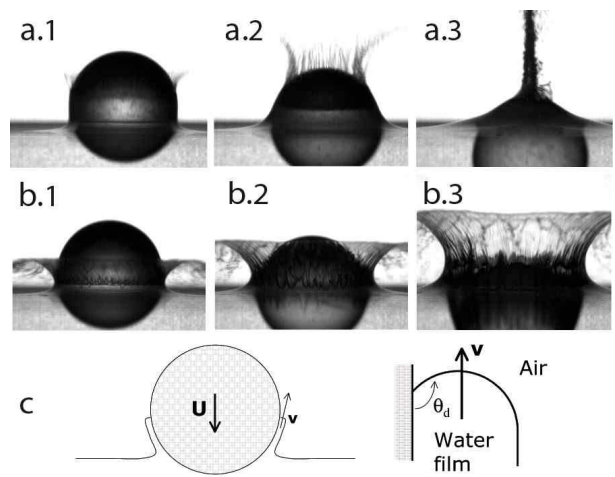}
\caption{Detailed chronophotography of impacting spheres with two different wettabilities at the same impact velocity $U=5$m.s$^{-1}$: (a.1)-(a.3) hydrophilic sphere (1.4, 2.2 and 3.9 ms after impact start); (b.1)-(b.3) hydrophobic sphere (1.5, 2.4 and 4.0 ms after impact start). For the hydrophilic sphere, the
considered impact velocity
is below the threshold for air entrainment: the ascending film is shown to follow the sphere and gather at the pole. For the hydrophobic sphere, the impact velocity is above the threshold for air entrainment : the ascending film detaches from the sphere, thereby creating a cavity during the impact.
(c) {\it left} - Sketch of the impact geometry; {\it right} - magnification of the triple line region. $\theta_d$ is the
dynamic contact angle, which is larger than the static contact angle $\theta_0$ for a moving triple line (with velocity $v$). The threshold velocity is reached as the contact line is no longer stable, which
occurs as $\theta_d\rightarrow180^\circ$.
\label{fig3}}
\end{center} 
\end{figure}
To rationalize these results, we zoom in on the detailed dynamics of the impact. An essential
characteristic of solid to liquid impacts is that a thin film develops during the impact and climbs up the impacting body \cite{Korobkin}. This film is evident in Fig.\ref{fig3} (a.1)-(a.3). However, the film
dynamics is seen to strongly differ depending on whether the velocity is below or above the threshold $U^{\star}$ for
air entrainement.  For the velocity considered in Fig.\ref{fig3}, the hydrophilic sphere is below the threshold: the film is seen to follow the sphere %up to the end of the impact, 
and closes up at the pole of the sphere (Fig.\ref{fig3} (a.1)-(a.3)). As such, no cavity is created. On the contrary, for the same velocity, the hydrophobic sphere is above threshold  and
the film is seen to
detach from the sphere {\it before} reaching the pole (Fig.\ref{fig3} (b.1)-(b.3)). The opened aperture left
at the top of the sphere then leads to cavity formation and air entrainment. 
These pictures thus point to the film dynamics
as the origin of air entrainment and splash. %raising the question of its stability.

We therefore propose an interpretation of these results in terms of contact line stability.
The geometry is described in Fig.\ref{fig3}(c). 
The liquid film and triple line move at a velocity $v$  of $v\approx \zeta U$, with  $\zeta \approx 2$ (see \cite{oliver}).  Let us first consider the motion of the film on a hydrophobic sphere. In this case the gas (air)
is the wetting phase and the solid surface moves {\it towards}Ê the non-wetting phase (liquid). 
This situation corresponds to the prototypical problem of forced (de)wetting, but here with the air replacing
the liquid in the role of the wetting phase, see Fig.\ref{fig3}c. Since the work of de Gennes \cite{deGennes}, it has been
known that
a critical speed exists above which the triple line is no longer stable, as the dynamic contact
angle $\theta_d$ goes to 180$^\circ$ \cite{Eggers}. Above this critical speed, the
solid will be coated by the wetting phase, here air. 
A hydrodynamic force balance at the contact line shows that this occurs
at a critical capillary number $Ca^\star=\mu v^\star/\gamma_{LV}$ obeying $Ca^\star\approx  \Theta_0^3/9\ell$, with $\Theta_0$ the static contact angle as defined with respect to the wetting phase (air) and $\ell\approx 15-20$, see \cite{deGennes,Eggers}.  Using $\Theta_0=\pi - \theta_0$, one gets
$Ca^\star \approx (\pi-\theta_0)^3/9\ell$. 
This classical reasoning, however, neglects dissipation in the non-wetting phase (here water).
This assumption is obviously not valid.  
We have added a liquid viscous contribution 
in the force balance at the triple line, in the form $F_L\approx C \mu_L v$ (with $C\sim 1$). This 
term adds to the classical contribution in the wetting phase corner, here $F_{air}(v)={3\mu_{air} \ell \over {\pi-\theta_d}} v$, diverging as $\theta_d\rightarrow 180^\circ$.
Both terms are of the same order of magnitude since $\alpha=3\ell \mu_{air}/\mu_{L}\sim 1$.  Using this new expression for the frictional force, the critical velocity is found to be of the form $v^\star= {g_0\over9\ell} {\gamma_{LV}\over \mu_{L}} [\pi-\theta_0]^{3}$, defining a critical
capillary number in terms of the {\it liquid} viscosity. The numerical prefactor $g_0$ 
is typically on the order of $\sim 5-10$, with a weak dependence on the liquid and gas viscosity. Using $v=\zeta U$, one  gets eventually the threshold velocity for non-wetting impacters ($\theta_0\geqslant 90^\circ$),
\begin{equation}
U^{\star}={g_0\over 9\ell\zeta} {\gamma_{LV}\over \mu_{L}}[\pi-\theta_0]^{3}.
\label{Unonwet}
\end{equation}
As shown in Fig.\ref{fig2}, this theoretical prediction is in very good agreement 
with the experimental results with hydrophobic impacters
($\theta\geqslant 90^\circ$). This mechanism culminates in
the superhydrophobic limit, for which the impacting body entrains air for any velocity.
Fixing $\ell=15$ and $\zeta=2$, experimental results are quantitatively reproduced with $g_0\approx 7$
(which corresponds to $C\simeq2.9$). 

The situation of a hydrophilic impacting sphere may be discussed along the same lines. While the
contact angle is lower than $90^\circ$ for small velocities, the dynamical contact angle $\theta_d$ will increase with the triple line velocity $v$.
As above, the triple line will disappear  as $\theta_d\rightarrow180^\circ$ \cite{Eggers}.
%, {\it i.e}  the dynamic contact angle measured with respect to the wetting phase approaches zero 
Unfortunately, no analytical description is available for $\theta_d(v)$ in this limit when starting from a wetting surface ($\theta_0\leqslant90^\circ$).
Nevertheless the physics are qualitatively similar to that described above for the non-wetting surfaces and
the dissipation in air, which diverges as $\theta_d\rightarrow 180^\circ$, will destabilize the contact line above a threshold velocity. 
We therefore expect again a critical velocity, scaling as in the previous case like 
\begin{equation}
U^\star \approx \xi {\gamma_{LV}\over \mu_L}.
\label{Uwet}
\end{equation}
The prefactor $\xi$ may 
depend on the static contact angle $\theta_0$. However, since at the threshold
for destabilization $\theta_d^\star\sim \pi$ is significantly larger than the static contact angle $\theta_0$, we only expect a weak dependence  of 
$\xi$ on $\theta_0$.
This point is confirmed experimentally, as shown in Fig.\ref{fig2}: in the wetting regime the
critical velocity for air entrainement is basically independent of the static contact angle. 
Moreover changing the fluid fully confirms 
the linear dependence of $U^\star$ on ${\gamma_{LV}\over \mu_L}$ as embodied in expression (\ref{Uwet}) (see Fig.\ref{fig2}, inset).
Comparison with experimental results suggests $\xi\approx0.1$ (see Fig.\ref{fig2}, inset).

We finish with a short discussion on the 'splashing' sound, which can be heard above threshold
(see Fig.\ref{fig0}).
The sound arises from the rapid closure of the cavity as it pinches off. These dynamics are gravity-driven
so that the closure time is typically $\tau\sim \sqrt{a/g}$, with $a$ the size of the impacting object and $g$ the gravitational constant \cite{Glasheen,Duclaux}. For a centimeter
body, $\tau\sim 100$ms, in full agreement with the recordings shown in Fig.\ref{fig0} (to compare
with the impact time $a/U\sim 5$ms in this case).  As we confirmed independently, 
the splash duration and amplitude  are therefore independent of the impact velocity! 

To conclude, our results give a new perspective on impacts in liquids, by pointing to the unexpected
role of surface properties. Air entrainement is best inhibited by using
clean and wetting surfaces. Hydrophobic objects make a splash.
%Making a big splashes require hydrophobic, big objects.
%wetting dynamics taking into account the gas phase : new

\vskip 2cm
{\bf \Large Methods}

%\begin{itemize}
{\bf Surface treatments to control wettability}\\
We use spheres made of glass, steel and aluminium, with diameters varying between  7 and 25.4mm.\\
%\begin{itemize}
%\item 
Wetting glass beads ($\theta_0\simeq 15-20^\circ$) are obtained by immersion during 40 min in a piranha solution [1 vol. H$_2$O$_2$, 2 vol. H$_2$SO$_4$], then rinsed using deionized water and isopropanol, and eventually heated at 110$^\circ$C for 20 min.\\
%\item 
Hydrophobic glass beads ($\theta_0\simeq 100-120^\circ$) are obtained by grafting silane chains on the surface. We chose
grafting of octyltriethoxysilane (105$^\circ$) or perfluorooctyltrichlorosilane ($110-120^\circ$) in the gas phase (by pumping in a closed vessel), for
15 hours at ambient temperature. After silanisation, the beads are rinsed with isopropanol, dried
and heated at 90$^\circ$ C during one hour. \\
%\item 
Superhydrophobic aluminium beads ($\theta_0\simeq
150-170^\circ$) are obtained following the chemical protocol proposed by Qian and Zhen \cite{Qian}.
The aluminium beads are first plunged in an aqueous solution of chlorhydric and fluorhydric acids 
for 15 s. Then a silane coating (perfluorooctyltriethoxysilane) is grafted on the beads
by silanisation in the liquid phase at ambient temperature for one hour and then heated at
130$^\circ$C for one hour.
This protocol works only for pure aluminium and we therefore used 1050 Al. 
%A superhydrophobic glass bead is obtained by coating the surface with a black carbon coating
%($\theta_0\simeq 170^\circ$). 
\\
%\item 
The contact angle on the steel beads is $\theta_0\sim 80-90^\circ$, obtained after 
cleaning with deionized water with detergent, and then with isopropanol.
%\end{itemize}
An AFM topographic scan of the beads' surface shows that for the beads considered in this study,
the peak-to-peak roughness was smaller than 100 nanometers (with a rms roughness of $\sim5$nm
for a $10\mu{\rm m}\times10\mu{\rm m}$ scan). Larger scale roughness was probed with a
profilometer (Tencor Instruments) with a $5\mu$m tip showing less than $20$nm rms deviation
over a millimeter scan.
For the superhydrophobic coatings, we have shown moreover
in a previous work using AFM measurements that the liquid interface on the coatings is very smooth with a peak-to-peak roughness in the hundreds of nanometers range \cite{Journet}. 

{\bf Impact experimental setup and protocole}\\
The beads are released from rest at varying heights above a transparent box containing the liquid.
The impact is recorded using a high speed video camera (Mikrotron) 
at a frame rate of $\sim1000$ fps. The impact speed is determined from the movie. 
Before each release, the beads are cleaned by rinsing with isopropanol, dried using azote and heated at
110$^\circ$C for 20 min for a complete drying. We left the beads cooling to ambient temperature
before impact.
Most of our experiments are conducted with water, but to study the effect of fluid characteristics
(viscosity, surface tension) we use
ethanol, isopropanol, and a water-glycerol mixture % with different fractions of glycerol 
(20wt\% of glycerol)
to vary the viscosity. 
The contact angle on the bare glass beads with these fluids was always smaller than $10^\circ$.
The viscosity of the water-glycerol mixture was measured before and after each impact using a Ubbelohde viscometer. %\NOTE{en cours de finalisation}.
Values for the surface tensions were taken from litterature.
%\end{itemize}

{\bf \small Acknowledgements} : This project was supported by the French Ministry of Defense, via DGA. L.B. would like to thank Jean-Fran\c cois Pinton for discussions on the subject and D. Huang for
a careful reading of the manuscript. 
The authors thanks S\'ebastien Manneville for kindly lending us the high speed video system
and Marie-Charlotte Audry for the AFM scans.

{\bf \small Conflict of interest} : the authors declare that they have no competing financial
interests.

{\bf \small Correspondance} and requests for materials should be addressed to L. B. (lyderic.bocquet@univ-lyon1.fr)

\end{document}